\documentclass[aps,twocolumn,nofootinbib]{revtex4}
\usepackage[latin1]{inputenc}
\usepackage{epsfig}
\usepackage{graphicx}
\usepackage{amsfonts}
\newcommand{\beq}{\begin{equation}}
\newcommand{\eeq}{\end{equation}}
\newcommand{\la}{\langle}
\newcommand{\ra}{\rangle}

\begin{document}

\title{Origins of the Ising model}
\author{Mário J. de Oliveira}
\affiliation{Universidade de São Paulo,
Instituto de Física,
Rua do Matão, 1371, 05508-090
São Paulo, SP, Brasil}

\begin{abstract}

In 1925, Ernest Ising published a paper analyzing a model
proposed in 1920 by Wilhelm Lenz for ferromagnetism.
The model is composed of constituent units that take only
two states and interact only when they are neighbors.
Ising showed  that in a linear chain the model does not
present an ordered ferromagnetic state, a frustrating but
correct result. However, Rudolf Peierls demonstrated in 1936
that the model does in fact present an ordered state in two
dimensions, and therefore in three dimensions. This result
reveals that short-range interaction and only two states for
each constituent unit are sufficient for ordering to occur
over long distances. These two elements are the key to
understanding the success of the model and its variants
even a hundred years after its appearance. Here we analyze
the emergence of the model in the period up to 1936.

\end{abstract}

\maketitle

\section{Introduction}

The Ising model
\cite{domb1960,brout1965,brush1967,ziman1972,salinas1997,niss2005}
was conceived by Wilhelm Lenz in 1920 \cite{lenz1920} and proposed
to his student Ernst Ising as a model for ferromagnetism. It was
analyzed by Ising in his doctoral thesis entitled {\it Contribution
to the Theory of Ferromagnetism}, defended at the University of
Hamburg. The results of the thesis were published the following year
in a paper of 1925 \cite{ising1925}. Although it is not a model for
every type of ferromagnetism, the model is reasonably appropriate
for the ferromagnetism of spin-half anisotropic magnetic systems
\cite{domb1964}.

The Ising model consists of interacting dipoles residing on the
sites of a lattice. The basic ingredients of the model, which
result in ferromagnetism, are the interaction only between
neighboring dipoles and only two states for each dipole. However,
these two ingredients are not inherent to magnetic systems and can
therefore be used in models to describe the ordering of systems of
other natures, such as the ordering in metallic alloys. More generally,
the model is capable of explaining and describing the {\it cooperative
state} of systems composed of interacting elements \cite{domb1960}.

Ising solved the one-dimensional model in the presence of a magnetic
field. The results showed that the magnetization vanishes when the
field vanishes, leading Ising to correctly conclude that the 
one-dimensional model does not describe the ferromagnetic state.
He further argued that the three-dimensional model would not exhibit
the ferromagnetic state, a frustrating result given the purpose of
the model. However, Herzfeld in 1925 \cite{herzfeld1925} referring to
Ising paper stated that the result in three dimensions would
need to be proven.

Lenz motivation lies in Pierre Weiss theory of ferromagnetism,
published in 1907
\cite{weiss1907}. In his exposition of the theory, Weiss shows that
the interaction between elementary magnetic dipoles cannot be magnetic
in nature, a result also commented on by Lenz. The interaction does
indeed have a quantum origin and was incorporated into the model for
ferromagnetism introduced by Heisenberg in a paper of 1928
\cite{heisenberg1928}. In this papaer, Heisenberg comments on
Ising result regarding the absence of ferromagnetism in a chain,
and, in a letter to Pauli dated 1928, he suggested that if Ising had
used a sufficient number of near neighbors, he would have obtained
ferromagnetism \cite{niss2005}.

In a lecture on magnetism delivered at the 1930 Solvay conference,
published in 1932 \cite{pauli1932}, Pauli used Heisenberg theory
in conjunction with Weiss molecular field to obtain the ferromagnetic
state. Pauli also obtains magnetization at low temperatures and
comments that the three-dimensionality of the crystal lattice is
essential for the emergence of ferromagnetism. He mentions the Ising
result on the absence of ferromagnetism in a chain but states that
this also occurs with quantum models. He adds that it is very likely
that an extension of Ising theory to the case of a three-dimensional
lattice would produce ferromagnetism.

In his book of 1932 \cite{vanvleck1932} on the theory of electric and
magnetic susceptibilities, van Vleck discusses Heisenberg theory
of ferromagnetism. He shows how to adapt Weiss molecular field
theory to obtain the ferromagnetic state and the critical temperature
of Heisenberg model. He claims that Ising results on the absence
of ferromagnetism are in apparent contradiction with his own,
justifying that this is supposedly due to Ising use of a coupling
between elementary magnetic dipoles as being the product of two
scalars rather than the scalar product of two vectors.

With the introduction of the Heisenberg model, considered more
appropriate for describing ferromagnetism, we should have expected
the Ising model to be forgotten \cite{brush1967,niss2005}.
However, the model reappeared as models for the ordering of
alloys and for the adsorption of gases on solid surfaces. 

Gorsky in 1928 \cite{gorsky1928}, Bragg and
Williams in 1934 \cite{bragg1934,bragg1935}, and Bethe in 1935
\cite{bethe1935} analyzed models for ordering in alloys that are
equivalent to the Ising model, although Ising is not cited.
Bethe paper is particularly interesting for two reasons. One
is his concern to show that the model he analyzes does not support
ordering in one dimension but does have ordering in two dimensions.
The other is the use of an approximate method that becomes exact in
one dimension. 

The model analyzed by Ising was cited by Fowler in his paper of 1935 
on the adsorption of gases on solid surfaces \cite{fowler1936}.
Fowler comments that the problem he analyzes was solved exactly
by Ising in a linear chain but not in two or more dimensions.
A method equivalent to Bethe was used by Peierls in
1936 \cite{peierls1936a} to analyze the Fowler model for adsorption.

In another paper of 1936 \cite{peierls1936}, Peierls acknowledges that
the models analyzed by Bragg and Williams, Fowler, and Bethe are
equivalent to the model analyzed by Ising. In this paper, whose title
contains the term 'Ising's model', Peierls demonstrates that in two
dimensions the Ising model does indeed exhibit the ordered state at
sufficiently low temperatures. Peierls argument leads us to the
conclusion that the ordered state also occurs in three-dimensional
models.

With the recognition that the Ising model is equivalent
to models of ordering in alloys, 
it came to be understood as a model for cooperative
phenomena \cite{niss2005}. In this sense, the model was studied by
Kramers and Wannier in 1941 \cite{kramers1941} as a model of statistical
mechanics. In this paper, they showed that the partition function,
and therefore the free energy, is related to the largest eigenvalue
of a certain matrix. Furthermore, they managed to determine the exact
value of the critical temperature in the square lattice through a
symmetry relation between high and low temperatures.

The calculation of the free energy of the two-dimensional model
in the absence of a field was performed by Onsager and published
in 1944 \cite{onsager1944}. The calculation showed that the free
energy $f$ as a function of temperature has a singularity at
the critical temperature of the type
$f-f_0\sim \varepsilon \ln |\varepsilon|$ where $\varepsilon$
is the deviation of the temperature from its critical value.
From this result, the following singular behavior is obtained
for the specific heat $c\sim \ln |\varepsilon|$. As for magnetization,
it cannot be determined from the free energy at zero field. However,
it can be obtained from correlations, a method used by Yang
\cite{yang1952} in 1952. Near the critical temperature,
he showed that magnetization behaves as $m\sim |\varepsilon|^{1/8}$.

Onsager and Yang results show that the critical behavior
represented by critical exponents can be distinct from those
obtained using Weiss theory \cite{oliveira2012}. For example,
from this theory, the critical exponent associated with
magnetization is $1/2$, while Yang result gives us the
value $1/8$. These results, and others such as spontaneous
symmetry breaking, explain the enormous success of the model
in the area of phase transitions and critical phenomena,
whose developement occurred mainly since the 1960s.

The Ising model, like other statistical mechanics models, is
defined not only by the interaction energy between the numerous
constituent units but also, and primarily, by the Gibbs probability
distribution, usually used implicitly. Since the Gibbs distribution
is the one that should occur in thermodynamic equilibrium, the
Ising model is understood as a model that describes a system in
thermodynamic equilibrium.

Thermodynamic equilibrium can also be understood as a final state
of a stochastic dynamic. In 1963, Glauber introduced a dynamic of
this nature constructed in such a way that the final distribution
is the Gibbs probability distribution associated with the Ising
model \cite{glauber1963}. This dynamic is defined by transition
rates that obey the detailed balancing condition, which ensures
that the distribution is equilibrium in the long run.

Stochastic dynamics \cite{tome2014} that do not obey detailed
balance lead the system in the long run to a steady state out of
thermodynamic equilibrium. A model of this type with transition
rates containing the two basic ingredients of the Ising model was
studied by the author in 1991 \cite{oliveira1992}. The study
revealed that such a model has an ordered state like the Ising
model and that the critical behavior is described by critical
exponents identical to those of the Ising model. These results
show that thermodynamic equilibrium is not a necessary condition
for the occurrence of the ordered state, reaffirming that the
ordered state is a consequence of only the two basic ingredients
mentioned. This is particularly relevant for the study of systems
that are active only when they are outside of thermodynamic
equilibrium, as is the case with biological systems.

\section{Weiss}

The theory of ferromagnetic ordering was not established by the
introduction of the Ising model. Pierre Weiss proposed in 1907 a
theory in which the ferromagnetic state is explained in terms of
a {\it molecular field} \cite{weiss1907}. In a paramagnetic system,
magnetization is null but can arise through an external field applied
to the system. Weiss argues that in a ferromagnetic system, magnetization
arises due to the field generated by the constituent units themselves,
the molecules of the system, and therefore he called it the molecular
field. More precisely, the field due to the neighbors of a given dipole
is responsible for the dipole orientation. Weiss consistently assumed
that the molecular field is proportional to the magnetization, so that
in the absence of magnetization, there is no molecular field. It is
important to note that Weiss emphasized that the molecular field arises
from non-magnetic forces.

In presenting the theory, Weiss begins with the Langevin formula
for the magnetization $m$ of a paramagnetic system subjected to a
magnetic field $H$. We write the Langevin formula as
\beq
m = \mu (\coth\alpha - \frac1\alpha), \qquad \alpha=\frac{\mu H}{kT},
\label{8}
\eeq
where $\mu$ is the magnetic moment of a molecule, $k$ is Boltzmann 
constant, and $T$ is the absolute temperature. The function of
$\alpha$ in parentheses on the right-hand side of equation (\ref{8})
grows monotonically with $\alpha$ and saturates asymptotically at 
the value 1. Weiss replaces $H$ with the molecular field $\lambda m$,
proportional to $m$, which is equivalent to writing
\beq
\alpha = \frac{\mu \lambda m}{kT}.
\label{10}
\eeq
One solution of equation (\ref{8}) with $\alpha$ given by (\ref{10})
is $m=0$, but there is another non-zero one that occurs at
temperatures lower than a critical temperature.

\begin{figure}
\begin{center}
\epsfig{file=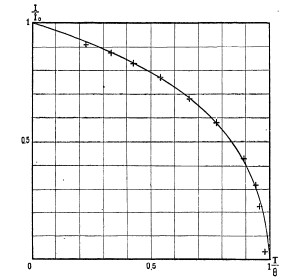,width=9cm}
\end{center}
\caption{\rm Figure from Weiss paper \cite{weiss1907} showing the
reduced magnetization $I/I_0$ as a function of the reduced temperature
$T/\theta$. The solid line represents the solution of equation
(\ref{15}). The plus signs represent experimental data obtained by
Weiss on a magnetite sample between $-79$ $^\circ$C and $587$ $^\circ$C.}
\label{weissfig}
\end{figure}

Next we obtain the non-zero solution for the case where $m$ is small
and therefore $\alpha$ is small. In this case the function in
parentheses in (\ref{8}) is $\alpha(1-c\alpha^2)/3$, where $c$ is a
positive numerical constant, and therefore equation (\ref{8}) reduces to
\beq
\frac{kT}{\mu\lambda}\alpha = \frac13\alpha(1- c\alpha^2),
\eeq
and we see that one solution is $\alpha=0$. The non-zero solution
is given by
\beq
c\alpha^2 = 1-\frac{3kT}{\mu\lambda},
\eeq
and exists if $T < \mu\lambda/3k$. The non-trivial solution
corresponds to spontaneous magnetization and therefore it
vanishes when $T = \mu\lambda/3k$, which is the critical
temperature.

Using Weiss original notation, the equation for magnetization,
which is equation (\ref{8}) with $\alpha$ given by (\ref{10}),
reads
\beq
\frac{I}{I_0} = \coth\alpha - \frac1\alpha,
\qquad \alpha = \frac{3I}{I_0}\frac{\theta}{T},
\label{15}
\eeq
where $I$ is the magnetization, $I_0$ is the saturation magnetization,
and $\theta$ is the critical temperature, called by Weiss the
temperature of the loss of spontaneous ferromagnetism. The solution
to this equation is shown in figure \ref{weissfig}.
Weiss also performed experiments to verify formula (\ref{15}) using
a sample of magnetite. Weiss experimental method consisted of shaping
the sample into an ellipsoid and suspending it by means of a torsion
spring in the presence of an electromagnet field. In equilibrium,
the magnetic torque is equal to that exerted by the spring and is
proportional to the square of the magnetization intensity. Weiss
experimental measurements were in the temperature range between that
of dry ice, the solid form of carbon dioxide, ($-79$ $^\circ$C) and
that of the disappearance of ferromagnetism in magnetite
($587$ $^\circ$C). The experimental values are presented
in figure \ref{weissfig}.

\section{Lenz}

The genesis of the Ising model is found in Lenz paper of 1920
\cite{lenz1920} on the magnetism of paramagnetic salts. Lenz argues
that elementary magnetic dipoles in crystalline solids are free to
assume various positions. A quantum treatment, however, shows that 
certain positions must be distinguished, for example, the extreme
positions. Assuming that the potential energy is large between the
two extreme positions, then only these should occur equivalently.
In the presence of an external magnetic field $H$, the equivalence
between these two positions disappears, resulting in a non-zero magnetic
moment. The probability of the occurrence of the position such that the
elementary dipole moment $\mu$ is in the direction of the field is
proportional to $e^\alpha$ where $\alpha=\mu H/kT$ and $T$ is the
absolute temperature. The probability of the other position, a dipole
in oppositition to the field, is smaller and proportional to
$e^{-\alpha}$. From these results,
Lenz obtains the average dipole moment,
\beq
\bar{\mu} = \mu \frac{e^{a}-e^{-a}}{e^{a} + e^{-a}}.
\eeq

Lenz then shows how spontaneous magnetization arises in ferromagnetic
bodies. He argues that the potential energy between neighboring
elementary magnetic dipoles favors one of two extreme positions,
thus leading to the emergence of spontaneous magnetization. He further
adds that ferromagnetic properties can be explained in terms of
non-magnetic forces, in agreement with Weiss view, who established
that the force between elementary dipoles is non-magnetic in nature.

\section{Ising}

\subsection*{Definition of the model}

Consider a sequence of $N$ sites along a straight line, equally
spaced. The sites are labeled from $1$ to $N$. To each site $i$
we associate a dipole that takes only two values along a given
direction. The two values are $+\mu$ and $-\mu$. Defining for
each site $i$ a variable $\sigma_i$ that takes the values $+1$
or $-1$, then the dipole of site $i$ can be written as
$\mu \sigma_i$. A state of the system is defined by the vector
$\sigma=(\sigma_1,\sigma_2,\ldots,\sigma_N)$. The total
magnetization of the system in a given state is therefore
\beq
m(\sigma) = \mu\sum_i \sigma_i.
\eeq

According to Lenz, only neighboring dipoles interact. Therefore,
the contribution to the total energy comes only from neighboring
dipoles. When they are parallel, the energy is lower than in the
antiparallel configuration. Considering the dipoles located at
neighboring sites $i$ and $i+1$, then they will be parallel when
$(\sigma_i,\sigma_{i+1})$ is equal to $(+,+)$ or $(-,-)$. The
antiparallel configuration occurs when $(\sigma_i,\sigma_{i+1})$
is equal to $(+,-)$ or $(-,+)$. In the parallel configuration,
the energy is assumed to be $-J<0$ and in the parallel condition
as being $J>0$. Thus, the energy between two neighboring sites can
be written as $-J\sigma_i\sigma_{i+1}$.

The total energy associated with a state $\sigma$ contains two
contributions. One is the sum of the interaction energies between
neighboring sites. The other is the contribution due to an external
field $H>0$. If the dipole located at $i$ is parallel to the field,
$\sigma_i=+1$, the contribution is $-H\mu$. If it is antiparallel
to the field, $\sigma_i=-1$, then the contribution is $H\mu>0$. In
both cases the contribution is $-H\mu\sigma_i$. The total energy
$E(\sigma)$ corresponding to a state $\sigma$ is therefore
\beq
E(\sigma) = -J\sum_{i=1}^{N-1} \sigma_i\sigma_{i+1}
- H\mu \sum_{i=1}^N \sigma_i.
\label{9}
\eeq

Assigning an energy to each state is not sufficient to determine
which state should occur. To do this, we assume that the different
states occur with a certain probability, which we choose as the one
introduced by Gibbs, which describes systems in thermodynamic
equilibrium, and is given by
\beq
P(\sigma) = \frac1Z e^{-\beta E(\sigma)},
\label{13}
\eeq
where $\beta=1/kT$, $k$ is Boltzmann constant and $T$ is the
absolute temperature, and $Z$ is a normalization factor such that
\beq
\sum_{\sigma_1}\sum_{\sigma_2}\ldots \sum_{\sigma_N} P(\sigma) = 1,
\eeq
which we write abbreviated in the following form
\beq
\sum_\sigma P(\sigma) = 1,
\eeq
and therefore $Z$ is given by
\beq
Z = \sum_\sigma e^{-\beta E(\sigma)}.
\eeq
Therefore, the Ising model is defined not only by the energy
function (\ref{9}), as is usually said, but also by the
probability distribution.

To determine the average magnetization, defined by
\beq
M = \sum_\sigma m(\sigma) P(\sigma),
\eeq
we use the following results. Differentiating
$Z$ with respect to $H$, we obtain
\beq
\frac{\partial Z}{\partial H}
= \beta \sum_\sigma m(\sigma) e^{-\beta E(\sigma)}.
\eeq
Dividing both sides by $Z$, and using (\ref{13})
\beq
\frac{\partial}{\partial H}\ln Z
= \beta \sum_\sigma m(\sigma) P(\sigma),
\eeq
and therefore
\beq
M =\frac1{\beta}\frac{\partial}{\partial H}\ln Z.
\eeq
We therefore see that the calculation of $M$ is reduced to
the determination of $Z$, known as the partition function.

\subsection*{Calculation of the partition function}

It is convenient to define
\beq
E_n (\sigma) = -J \sum_{i=1}^{n-1} \sigma_i\sigma_{i+1}
- H\mu \sum_{i=1}^n \sigma_i,
\eeq
and the quantity
\beq
Z_n(\sigma_n) = \sum_{\sigma_{n-1}}\ldots\sum_{\sigma_1}
e^{-\beta E_n(\sigma)},
\eeq
and note that we are not summing in $\sigma_n$. Using these
definitions we obtain the following relation
\beq
Z_{n+1}(\sigma_{n+1}) = \sum_{\sigma_n}
e^{\beta J\sigma_{n+1}\sigma_n+\beta H\sigma_{n+1}}
Z_{n}(\sigma_n).
\eeq
Using the notation $A_n=Z_n(+1)$ and $B_n=Z_n(-1)$, then
\beq
A_{n+1} =
e^{\beta J+\beta H} A_{n}
+ e^{-\beta J+\beta H} B_{n},
\eeq
\beq
B_{n+1} =
e^{-\beta J-\beta H} A_{n}
+ e^{\beta J -\beta H} B_{n},
\eeq

Assuming solutions of the type
\beq
A_n = a \lambda^n, \qquad B_n = b\lambda^n,
\eeq
then
\beq
e^{\beta J+\beta H} a
+ e^{-\beta J+\beta H} b = \lambda a,
\eeq
\beq
e^{-\beta J-\beta H} a
+ e^{\beta J-\beta H} b = \lambda b,
\eeq
which is an eigenvalue equation. Eigenvalues are solutions of
\beq
(e^{\beta J+\beta H} - \lambda)
(e^{\beta J-\beta H} - \lambda)
- e^{-2\beta J} = 0,
\eeq
that is, they are solutions of the equation
\beq
\lambda^2 - 2 e^{\beta J}(\cosh\beta H) \lambda
+ e^{2\beta J} - e^{-2\beta J} = 0.
\eeq
The solutions are
\beq
\lambda_1 = e^{\beta J}(\cosh\beta H) 
+e^{\beta J}\sqrt{(\sinh\beta H)^2
+ e^{-4\beta J} },
\eeq
\beq
\lambda_2 = e^{\beta J}(\cosh\beta H) 
- e^{\beta J}\sqrt{(\sinh\beta H)^2
+ e^{-4\beta J} }.
\eeq
The general solution is then
\beq
A_n = a_1 \lambda_1^n + a_2 \lambda_2^n, \qquad
B_n = b_1 \lambda_1^n + b_2 \lambda_2^n.
\eeq
Since the partiion function is $Z_N=A_N + B_N$ then 
\beq
Z = c_1\lambda_1^N + c_2 \lambda_2^N.
\eeq

\subsection*{Magnetization}

The magnetization per site $m=M/N$ is obtained
through
\beq
m = \beta \frac{\partial}{\partial H} \frac1N\ln Z,
\eeq
and we will do this calculation for very large $N$.
Using the previous result for $Z$, we can write
\beq
\frac1N\ln Z = \ln \lambda_1 +
\frac1N\ln (c_1 + c_2 \frac{\lambda_2^N}{\lambda_1^N}).
\eeq
Now we note that $\lambda_1>\lambda_2$ for $T\neq0$
and therefore if $N$ is large enough the fraction
inside the parentheses can be neglected. Therefore
the magnetization per site is given by
\beq
m = \beta \frac{\partial}{\partial H} \ln\lambda_1,
\eeq
and is therefore related to the larger of the two eigenvalues.

Differentiating $\ln\lambda_1$ with respect to $H$ we obtain
\beq
m = \frac{\sinh\beta H}{\sqrt{(\sinh\beta H)^2
+ e^{-4\beta J}}},
\eeq
which is the result obtained by Ising \cite{ising1925} and
contained in his 1925 paper. When $H\to0$ we see that
$m\to 0$ and therefore there is no spontaneous magnetization
in the one-dimensional Ising model.

\begin{figure}
\begin{center}
\epsfig{file=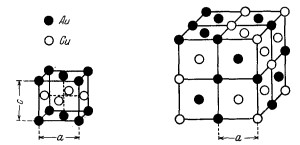,width=7cm}
\end{center}
\caption{\rm Figure from Gorsky paper \cite{gorsky1928}
illustrating the ordered state, on the left, and the disordered
state, on the right, of the AuCu alloy.}
\label{gorskyfig}
\end{figure}

\section{Gorsky}

The ferromagnetic state can be understood as a state with magnetic
order. We can imagine that systems of other natures could also
exhibit ordered states. This indeed occurs, for example, in metallic
alloys. The occurrence of the ordered state in metallic alloys was
predicted by Tammann in 1919 and subsequently confirmed by other
authors in various alloys. In a paper of
1928 on the CuAu alloy, Vadim Gorsky \cite{gorsky1928}
introduced a statistical model to describe the transition from
the disordered to the ordered state in this crystalline alloy.

A crystalline alloy is a solid composed of two or more types of atoms
arranged in a crystalline structure. Typically, each site in the
crystal lattice can be occupied by any type of atom, and the alloy
is understood as a homogeneous mixture. However, it is possible for
atoms of one type to be preferentially located in certain sites of
the crystal lattice, as illustrated in figure \ref{gorskyfig}.
Consider a binary alloy in a crystal lattice consisting of two
intertwined sublattices and assume that half of the atoms are type
A and the other half are type B. If the atoms of type A are
preferentially located in one sublattice and therefore the atoms
of type B preferentially in the other, then we are dealing with an
ordered state. The ordered state disappears when the atoms A and B
are equally distributed in either of the two sublattices. This state
is called disordered and should not be confused with the disordered
state that characterizes amorphous solids.

To understand Gorsky model, we consider the crystal lattice just
described, composed of two sublattices labeled 1 and 2. The lattice
contains $N$ atoms, half of which are type A and the other half are
type B. The degree of ordering is characterized by the number of
atoms of each type in one of the sublattices, say sublattice 1,
since the other will have complementary numbers. Furthermore, it
is sufficient to refer to the number of atoms $n$ of one type,
say type A, since the number of atoms of another type will be $N-n$.
Gorsky chooses the fraction $\alpha=n/N$ to represent the degree
of ordering. Ordering occurs for $1\geq\alpha>1/2$, and $\alpha=1/2$
characterizes the disordered state.

In his model, Gorsky refers to ``ordered'' sites and ``disordered''
sites which we interpret as the sites of sublattice 1 and sublattice 2,
respectively, relative to the atoms of type A. Ordering is reduced
to determining the number of atoms in the ``ordered'' sites, which
means determining the number of atoms of type A in sublattice 1,
which we denote by $n$.

The Gorsky model is a dynamic model and corresponds to a stochastic
process. Following Gorsky, we denote by $a_1$ the rate of transition
of an atom A from 1 to 2 and by $a_2$ that of an
atom A from 2 to 1. These rates depend on temperature. The number
of A atoms that are transferred per unit time from 1 to 2 is
therefore $n a_1$ and from 2 to 1 is $(N-n)a_2$.
In equilibrium we must have
\beq
n a_1 = (N-n)a_2,
\eeq
and therefore
\beq
\alpha = \frac1{1+ a_1/a_2}.
\label{25}
\eeq

Assuming that in the equilibrium state the probability distribution
of the configurations is in agreement with the Gibbs equilibrium
distribution then the rates must obey the condition
\beq
a_1 e^{-u_1/kT}=a_2 e^{-u_2/kT},
\eeq
where $u_1$ and $u_2$ are the energies of A at a
site on sublattices 1 and 2, respectively. Therefore
\beq
\frac{a_1}{a_2} = e^{-(u_2-u_1)/kT}.
\eeq
Gorsky introduces the hypothesis that as the system approaches
the disordered state, that is, as $\alpha$ approaches $1/2$,
the difference $u_2-u_1$ vanishes.
He then admits the following relation
\beq
u_2-u_1= c(2\alpha -1),
\label{26}
\eeq
and therefore
\beq
\frac{a_1}{a_2}= e^{-c(2\alpha -1)/kT}.
\eeq
Substituting this result into (\ref{25}), Gorsky
arrives at the following relation
\beq
\alpha = \frac1{1+ e^{-c(2\alpha -1)/kT}},
\label{30}
\eeq
which is an equation that determines $\alpha$. This equation 
has a solution $\alpha=1/2$ that describes the disordered state.
But it also has a solution such that $\alpha>1/2$ that
describes the ordered state and occurs if $T<c/2k$.

It is worth noting the analogy of this model with Weiss
molecular field theory. Gorsky hypothesis (\ref{26}) is
analogous to Weiss hypothesis that the molecular field
is proportional to the magnetization. If we use the
parameter $m$, defined by $m=2\alpha-1$, then equation
(\ref{30}) can be written in the form
\beq
m = \tanh \frac{c\,m}{2kT},
\label{31}
\eeq
which is analogous to equation (\ref{8}) supplemented
by equation (\ref{10}).

\section{Bragg and Williams}

In a 1934 paper \cite{bragg1934}, William L. Bragg and Evan Williams
proposed a model for the ordering of alloys that is equivalent to
the model proposed by Gorsky. However, they extended the model to
more complex lattices than the one considered by Gorsky. In a
subsequent paper of 1935 \cite{bragg1935}, they acknowledged
Gorsky work. Below, we describe the Bragg and Williams model
in its simplest form, in which the crystal lattice consists of
two sublattices denoted by $\alpha$ and $\beta$ and by two types
of atoms, A and B. With respect to the A atoms, the sites of the
$\alpha$ sublattice are understood as positions of order and those
of the $\beta$ sublattice as positions of disorder. Denoting by $p$
the probability of a site $\alpha$ being occupied by an atom A then
the degree of order is defined by $S=2p-1$, and $0<S\leq1$
and $1/2<p\leq1$, in the ordered state, and $S=0$ and $p=1/2$
in the disordered state.

Assuming initially that the atoms do not interact, then the
potential energy associated with a site depends only on which
atom occupies the site. Let $V_{\rm a}$ and $V_{\rm b}$ be the
potential energies associated with a site $\alpha$ when it is
occupied by an atom A and B, respectively. Using the Boltzmann
relation, the probabilities of occupation by an atom A and B are
respectively proportional to $e^{-V_{\rm a}/kT}$ and
$e^{-V_{\rm b}/kT}$. The ratio $p/(1-p)$ between
these probabilities is therefore
\beq
\frac{p}{1-p} = e^{- V/kT},
\eeq
where $V=(V_{\rm a}-V_{\rm b})/2$. From this relation we obtain
\beq
p = \frac{1}{1+e^{V/2kT}},
\eeq
and from it we find $S=2p-1$,
\beq
S = \tanh \frac{V}{4kT}.
\label{32}
\eeq

Bragg and Williams then introduce the hypothesis that
$V$ is proportional to $S$,
\beq
V = V_0 S.
\eeq
This relationship is analogous to that used by Weiss that
the molecular field is proportional to the magnetization
and also used by Gorsky, as we saw above. From this
hypothesis, equation (\ref{32}) becomes the equation
\beq
S = \tanh \frac{V_0 S}{4kT},
\label{33}
\eeq
and we see that it is equivalent to equation (\ref{30})
obtained by Gorsky. To do this, simply compare it with
(\ref{31}) which is equivalent to equation (\ref{30}).

\begin{figure}
\begin{center}
\epsfig{file=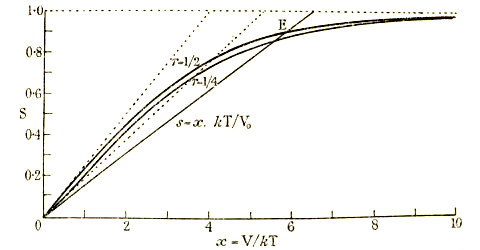,width=8.5cm}
\end{center}
\caption{\rm Figure from the paper by Bragge and Williams
\cite{bragg1934} showing the graphical solution of equation
(\ref{33}), upper continuous curve ($r=1/2$), and of equation
(\ref{33a}) for $r=1/4$, lower continuous curve.}
\label{braggfiga}
\end{figure}

Equation (\ref{33}) has a solution $S=0$ and also a non-zero
solution as long as the temperature is less than the critical
temperature. To show the existence of a non-zero solution,
Bragg and William use a graph showing the functions
$f=\tanh x/4$ and $g=kTx/V_0$, as shown in figure
\ref{braggfiga}. The intersection of these two functions
gives the solution to equation (\ref{33}). When $x$ is small,
we can use the approximation $f=x/4$. Therefore, if $kT/V_0>1/4$,
the only solution is $x=0$, or $S=0$. For $kT/V_0<1/4$, the
non-zero solution appears, and therefore the critical
temperature is given by $kT_c/V_0=1/4$, or
\beq
T_c = \frac{V_0}{4k}.
\eeq

\begin{figure}
\begin{center}
\epsfig{file=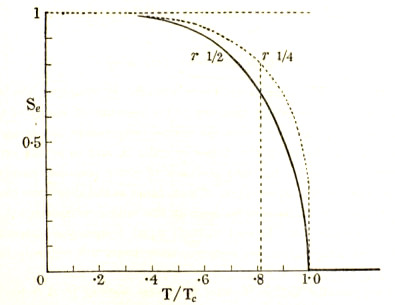,width=7cm}
\end{center}
\caption{\rm Figure from the paper by Bragg and Williams 
\cite{bragg1934} showing $S$ as a function of temperature
corresponding to the solution of equation (\ref{33}), upper
solid curve ($r=1/2$), and of equation (\ref{33a})
for $r=1/4$, dotted curve.}
\label{braggfigb}
\end{figure}

To explicitly determine the non-zero solution near the critical
temperature, simply expand $f$ to cubic order terms and compare
and equate to $g$. The result is
\beq
S - \frac13 S^3 = \frac{T}{T_c}S.
\eeq
The non-zero solution of this equation is given by
\beq
S^2  = 3(T_c-T)/T_c.
\eeq
The general solution of $S$ as a function of temperature
is shown in figure \ref{braggfigb}.

The results presented above correspond to the simplest model in
which the crystal lattice divides into two sublattices with the
same number of sites in each. Bragg and Williams also consider
more complex cases in which the crystal lattice divides into
more than two sublattices. Considering $n$ of sublattices,
then equation (\ref{33}) is replaced by
\beq
S = 1 - \frac{\{4r(1-r)(e^x-1)+1\}^{1/2}-1}
{2r(1-r)(e^x-1)}
\label{33a}
\eeq
where $r=1/n$.

\section{Bethe}

The theoretical approaches of Gorsky and of Bragg and Williams to
the ordering of atoms in metal alloys are similar to that employed
by Weiss for ferromagnetism. Hans Bethe treatment of this problem
on the other hand differs from these approaches and comes closer to
that given by Lenz and Ising for ferromagnetism although Bethe does
not name either of these two authors \cite{bethe1935}.

The model considered by Bethe describes a binary alloy AB consisting
of two sublattices. The sublattices are intertwined in such a way
that all neighboring sites of a site on one sublattice belong to
the other sublattice. The number of A atoms is equal to the number
of B atoms, and the completely ordered state is one in which A atoms
occupy the sites of one sublattice and B atoms occupy the other
sublattice. The atoms interact only when they are in neighboring sites.
The interaction energy between an A atom and a B atom is $V_{\rm ab}$,
between two A atoms is $V_{\rm aa}$, and between two B atoms is
$V_{\rm bb}$. The energy $V_{\rm ab}$ is lower than the other two,
so the completely ordered state has the lowest energy.

Bethe solves the model using a statistical method to approximately
determine the number of pairs of sites of the various types:
AA, BB, and AB. This method has the peculiarity of solving the
one-dimensional model exactly, concluding that there is no
ordered state. However, Bethe argues before using the method
that the interaction between first neighbors is not sufficient
to establish the ordered state in the one-dimensional model.

The critical temperature obtained by Bethe for a regular lattice
is proportional to the inverse of $\ln[z/(z-2)]$, where $z$ is
the number of neighbors of a site. In one dimension $z=2$ and
therefore the critical temperature is zero, which means that
there is no ordering at finite temperature and the disordered
state occurs at any finite temperature in one dimension.

We then derive Bethe results using a different approach from
the one originally employed by Bethe. The method we use is a
reformulation of that proposed by Ferreira, Salinas, and Oliveira
in 1977 \cite{ferreira1977}. To do this, we begin by representing the
configuration of the atoms by means of a variable $\sigma_i$ that
assumes the values $+1$ or $-1$ according to the following convention.
If site $i$ belongs to one of the sublattices, then $\sigma_i=+1$ if
the site is occupied by an atom A and $\sigma_i=-1$ if it is occupied 
y B. If site $i$ belongs to the other sublattice, the opposite occurs:
$\sigma_i=-1$ if the site is occupied by A and $\sigma_i=+1$ if the
site is occupied by B. Note that using these variables, the ordered
state is one in which all variables $\sigma_i$ have the same sign.

Using these variables the total energy of the system is
\beq
E = - J\sum_{(ij)}\sigma_i\sigma_j,
\eeq
where $J=(V_{aa} + V_{bb}- 2 V_{ab})/4>0$ and the sum extends
over pairs of neighboring sites. The probability $P(\sigma)$
of the occurrence of a configuration $\sigma$ is given by
the Gibbs equilibrium distribution
\beq
P(\sigma) = \frac1Z e^{-\beta E(\sigma)},
\label{28}
\eeq
where $\beta=1/kT$. 

From the Gibbs distribution we can determine several marginal
distributions. The one that interests us here is the one associated
with a central site labeled zero and its $z$ nearest neighbors,
which we denote by $P(\sigma_0,\sigma^{\rm v})$ where
$\sigma^{\rm v}=(\sigma_1,\ldots,\sigma_z)$ denotes the collection
of variables associated with the neighboring sites of the central
site. This distribution is obtained from the Gibbs distribution
by summing over all variables other than those associated with
the central site and its neighbors.

Next we derive an equation that $P(\sigma_0,\sigma^{\rm v})$ must
obey to ensure that it describes a system in equilibrium, that is,
that it is a marginal distribution of $P(\sigma)$ given by
(\ref{28}). To do this we start with the following identity
that follows directly from (\ref{28})
\beq
\frac{P(\sigma^i)}{P(\sigma)} =  \frac{e^{-\beta E(\sigma^i)}}
{e^{-\beta E(\sigma)}},
\eeq
which we write in the form
\beq
P(\sigma^i) = e^{-\beta [E(\sigma^i)-E(\sigma)]} P(\sigma).
\eeq
The term in brackets depends only on site $i$ and its neighbors.
Therefore, we can perform the sum on both sides of this equation 
in the variables of the other sites to obtain for $i=0$ the relation
\beq
P(-\sigma_0,\sigma^{\rm v}) = e^{-2\beta J \sigma_0 s}
P(\sigma_0,\sigma^{\rm v}),
\label{38}
\eeq
where $s$ is an abbreviation for the sum of the variables
associated with the neighbors of site 0. We call this
equation the thermodynamic equilibrium relation.

The approach we consider consists of using an approximation to
$P(\sigma_0,\sigma^{\rm v})$ obtained as follows. We first write 
\beq
P(\sigma_0,\sigma^{\rm v})
= P(\sigma^{\rm v}|\sigma_0)P(\sigma_0),
\eeq
and then we approximate the conditional probability distribution
by the product of conditional probabilities associated with
pairs of neighboring sites, that is,
\beq
P(\sigma^{\rm v}|\sigma_0)P(\sigma_0) = \prod_j P(\sigma_j|\sigma_0).
\eeq
Since
\beq
P(\sigma_j|\sigma_0) = \frac{P(\sigma_0,\sigma_j)}{P(\sigma_0)},
\eeq
then the approximation to $P(\sigma_0,\sigma^{\rm v})$ is given by
\beq
P(\sigma_0,\sigma^{\rm v}) = [P(\sigma_0)]^{-z+1}
\prod_j P(\sigma_0,\sigma_j).
\eeq

We assume that the pair distribution is of the form
\beq
P(\sigma_0,\sigma_j) = \frac1{Z_2}
e^{K \sigma_0\sigma_j + H_2(\sigma_0 + \sigma_j)},
\label{42a}
\eeq
where $K$ and $H_2$ are parameters to be determined and
\beq
Z_2 = 2e^{2K}\cosh 2H_2 + 2e^{-K}
\eeq 
As for the distribution $P(\sigma_0)$ it is given by
\beq
P(\sigma_0) = \frac1{Z_1} e^{H_1\sigma_0},
\label{42b}
\eeq
where $H_1$ is a parameter to be determined and
\beq
Z_1 = 2 \cosh H_1.
\eeq

Since the one-site distribution must be the marginal distribution
of the pairwise distribution, then $m=\la \sigma_0\ra$ must have
the same value whether we use the one-site distribution or the
pairwise distribution. Using the former, we obtain
\beq
m = \tanh H_1.
\label{45a}
\eeq
Using the second one, we get
\beq
m = \frac{e^{2K}\sinh 2H_2}{e^{2K}\cosh 2H_2 + 1}.
\label{45b}
\eeq

Now we use the thermodynamic equilibrium relation (\ref{38})
to find the relationship between the parameters $K$, $H_1$
and $H_2$. From (\ref{42a}) and (\ref{42b}), we obtain
\beq
\frac{P(-\sigma_0,\sigma^{\rm v})}{P(\sigma_0,\sigma^{\rm v})} =
e^{-2[zH_2-(z-1)H_1]\sigma_0 - 2K \sigma_0 s},
\eeq
where $s$ represents the sum of the variables associated with the
neighbors of the central site. Comparing with (\ref{38}), we
obtain the relations
\beq
K=\beta J, \qquad \frac{H_1}{z} = \frac{H_2}{z-1}.
\label{45c}
\eeq

Relations (\ref{45a}), (\ref{45b}), and (\ref{45c}) determine
$m$ as a function of temperature. One solution of these equations
is $m=0$, which corresponds to the disordered state. However,
there is a nonzero solution that corresponds to the ordered state
that occurs at a temperature lower than a critical temperature.
To determine it, we write equations (\ref{45a}) and (\ref{45b})
for small values of $H_1$ and $H_2$, which are
\beq
m = H_1, \qquad m = 2 H_2\,\frac{e^{2K}}{e^{2K} + 1}.
\eeq
Using (\ref{45c}) we obtain the relation
\beq
e^{2\beta J} = \frac{z}{z-2} 
\eeq
Remembering that $\beta=1/kT$, we obtain from this relationship
the critical temperature
\beq
\frac{2J}{kT_c} = \ln \frac{z}{z-2},
\eeq
which is the result obtained by Bethe.

\section{Fowler}

The work of Bragg and Williams and Bethe on order-disorder in alloys
inspired Ralph Fowler to develop a similar model for gas adsorption
on solid surfaces, published in a paper of 1936 \cite{fowler1936}.
Fowler turns his attention to adsorption by a single monomolecular
layer. He acknowledges that the existence of a critical temperature
in this system had been previously discussed by Frenkel in 1924
\cite{frenkel1924}.

Frenkel analyzes the formation of a layer of atoms adsorbed on a
surface. He assumes the following equation for the number of atoms
$n$ adsorbed on a surface,
\beq
\frac{dn}{dt} = \nu - \alpha n + \beta n^2.
\eeq
where $\nu$ is the rate of arrival of atoms at the surface, and
the other two terms refer to the departure of atoms from the
surface. The term $\alpha n$ refers to isolated atoms and the
term $\beta n^2$ refers to bonded atoms. In the steady state,
$dn/dt=0$, the equation to be solved is
$\beta n^2 - \alpha n + \nu = 0$, which has a solution provided
$\nu\leq\alpha^2/4\beta$. Therefore, there is a critical value
\beq
\nu_c = \frac{\alpha^2}{4\beta}
\eeq
above which the formation of the adsorbed layer becomes impossible.
Since $\alpha$ and $\beta$ depend on temperature, Frenkel concludes 
hat $\nu_c$ must correspond to a critical temperature.

In Fowler model, a solid surface contains $N$ sites where atoms
can be deposited. Each site can accommodate only one atom. Fowler
assumes that the energy of a layer of $M$ atoms is
\beq
E = - M h - X \varepsilon
\eeq
where $-h$ is the energy associated with an atom, $X$ is the number
of pairs of nearby neighbors, and $-\varepsilon$ is the interaction
energy associated with a pair of neighboring atoms.
The partition function is determined by
\beq
\zeta(M,T) = \sum_X g(M,X) e^{(Mh+X\varepsilon)/kT},
\eeq
where $g$ is the number of arrangements of $M$ atoms in $N$ sites
that produce $X$ pairs. Fowler claims that this problem
was solved exactly by Ising in a linear chain but not in
two-dimensional or higher-dimensional lattices.
 
The calculation of the function is calculated by an approximate
method that consists of assuming that $X=M\theta/2$, $\theta=M/N$
which Fowler claims is the same type of approximation used by
Bragg and Williams. With this approximation and considering that
\beq
\sum_X g(M,X) = {N\choose M},
\eeq
then
\beq
\frac1N\ln \zeta = - \theta\ln \theta
- (1-\theta)\ln (1-\theta)
+ \frac1{kT}(h\theta+ \frac\varepsilon2 \theta^2).
\eeq
From this equation Fowler obtains the result
\beq
\frac{\partial\ln\zeta}{\partial\theta}
= \ln \frac{1-\theta}{\theta}
+ \frac1{kT}(h+ \varepsilon \theta).
\label{17}
\eeq

To simplify Fowler results, we use the abbreviation
$\mu=-kT\partial\ln\zeta/\partial\theta$ to write
equation (\ref{17}) as
\beq
\mu = -kT \ln \frac{1-\theta}{\theta}
- h - \varepsilon \theta.
\label{17a}
\eeq
The quantity we abbreviate as $\mu$ is related to the pressure
and temperature of the gas that is in equilibrium with the
adsorbed layer. Therefore, $\mu$ is a function of pressure
and temperature, and equation (\ref{17a}) gives $\theta$
as a function of $\mu$.

\section{Peierls}

\subsection*{Adsorption model}

In 1936, Rudolf Peierls published two consecutive papers
related to the Ising model. In the first, \cite{peierls1936a},
he analyzes the model for the adsorption of atoms on solid
surfaces that had been introduced by Fowler and published in
the same year, 1936. Peierls uses an approximate method that
is equivalent to that introduced by Bethe.

The model and approach used by Peierls are as follows.
1) Each site in a regular lattice of $N$ sites can accommodate
a single atom. 2) The adsorption energy of each atom is $h$.
3) The interaction energy of each pair of neighboring atoms is
$-V$. Using a variable $\eta_i$ that takes the values $+1$ if
site $i$ is occupied and $0$ if site $i$ is empty, then the
number $M$ of sites occupied by an atom is
\beq
M = \sum_i \eta_i,
\eeq
and the number $X$ of pairs of neighboring sites occupied by atoms is
\beq
X = \sum_{(ij)}\eta_i\eta_j,
\eeq
in which the sum extends over pairs of neighboring sites.
The total energy of a configuration $\eta$ is
\beq
E = \mu M - V X,
\eeq
and the probability of a configuration of atoms is given by
\beq
P = \frac1Z e^{-E/kT}.
\eeq

Next, we consider the marginal probability distribution associated
with a central site $i=0$ and its $z$ neighboring sites.
The occupancy variables for these sites are 
$\eta_0,\eta_1,\ldots,\eta_z$.
For convenience we define
\beq
E_0 = \mu (\eta_0+\sigma_1+\ldots+\sigma_n)
- V\sigma_0(\sigma_1+\ldots+\sigma_n)
\label{16}
\eeq
and also $E'=E-E_0$. Therefore, $P(\sigma)$ can be written as
\beq
P = \frac1Z e^{-E_0/kT-VE'/kT}.
\eeq
We note that $E_0$ depends only on the variables of the central
site and its neighbors. Summing $P(\eta)$ over all variables
except these, we obtain
\beq
P(\eta_0,\eta_1,\ldots,\eta_z)
= \frac1Z e^{-E_0/kT} 
\phi(\eta_1,\ldots,\eta_z),
\eeq
where $\phi$ is the sum of $e^{-VE'/kT}$ in all variables except
the variables of the central site and its neighbors. We note
that $\phi$ does not depend on $\sigma_0$ because $E'$ does not
depend on that variable. Using (\ref{16}) we can write
\beq
e^{-E_0/kT}= a^{\eta_0} b^{\eta_0(\eta_1+\ldots+\eta_z)},
\eeq
where we are using the abbreviations $a=e^{-\mu/kT}$ and
$b=e^{V/kT}$. Then $\phi$ is approximated by
\beq
\phi = c^{\eta_1+\ldots+\eta_z},
\eeq
where $c$ is a constant to be determined, so that
\beq
P(\eta_0,\eta_1,\ldots,\eta_z)
= \frac1Z a^{\eta_0} b^{\eta_0(\eta_1+\ldots+\eta_z}
c^{\eta_1+\ldots+\eta_z}.
\label{18}
\eeq

Before proceeding, we need to determine $Z$, which is given by
\beq
Z = (1 + c)^z + a (1 + b c)^z.
\eeq
From the probability distribution (\ref{18}), we determine the
probabilities $P_0(\eta_0)$ and $P_1(\eta_1)$ associated with
the central site and one of its neighbors. Summing the
probability (\ref{18}) over all variables except $\eta_0$,
we obtain
\beq
P_0(\eta_0) = \frac1Z a^{\eta_0} (1+b^{\eta_0}c)^z.
\label{27}
\eeq
Summing the probability (\ref{18}) in all variables
except $\eta_1$, we obtain
\beq
P_1(\eta_1) = \frac1Z [c^{\eta_1}(1+c)^{z-1}
+ a (bc)^{\eta_1}(1+bc)^{z-1}].
\label{22}
\eeq

Since the system is homogeneous, the probability of occupation
of a site is independent of the site, which leads us to the
condition $P_0(1)=P_1(1)$. The fraction $\theta=M/N$ of sites
occupied by a molecule is therefore equal to $P_0(1)$.
Using (\ref{27}), we obtain the equation
\beq
\theta = \frac{a (1+bc)^z}{(1 + c)^z + a (1 + b c)^z}.
\label{34}
\eeq
Similarly, $\theta$ is equal to $P_1(1)$ and using (\ref{22}),
we obtain the equation
\beq
\theta = \frac{c(1+c)^{z-1} + abc(1+bc)^{z-1}}
{(1 + c)^z + a (1 + b c)^z}.
\eeq
Equating the equations we obtain
\beq
c = a (\frac{1+bc}{1 + c})^{z-1}.
\label{29}
\eeq

The solution to equation (\ref{29}) gives us $c$ in terms of
$a$ and $b$, which substituting into (\ref{34}) gives us
$\theta$. To determine the solution to these equations we
use the abbreviation
\beq
r = \frac{1+bc}{1 + c},
\eeq
and therefore $c=(r-1)(b-r)$. Equations (\ref{34}) and
(\ref{29}) can then be written as
\beq
2\theta-1= \frac{a r^z-1}{a r^z+1},
\label{35}
\eeq
\beq
a r^z - ab r^{z-1} + r= 1.
\label{36}
\eeq
Solving this equation we determine $r$, which substituted
into (\ref{35}) gives us $\theta$.

Next we consider the case where $a$ and $b$ are related by
$a^2 b^z=1$. Furthermore we consider the case where $ar^z$
is close to unity, which means that $\theta$ is close to $1/2$.
Defining the deviation $\varepsilon$ by $ar^z=1+\varepsilon$,
then up to terms of order $\varepsilon^3$, equation (\ref{36})
becomes
\beq
\frac1{\sqrt{b}}\,\varepsilon
- \frac{z-2}{z}\varepsilon 
+ \frac{(z-2)(z-1)}{6z^3}\varepsilon^3 = 0.
\eeq
A solution to this equation is $\varepsilon=0$ which
gives $\theta=1/2$. There is however a non-zero
solution to $\varepsilon$ given by
\beq
\frac{(z-2)(z-1)}{6z^3}\varepsilon^2 =\frac{z-2}{z}
- \frac1{\sqrt{b}},
\eeq
that exists if $z>2$ and for $\sqrt{b} > z/(z-2)$.
Remembering that $b=e^{V/kT}$ this condition gives us
\beq
\frac{V}{2kT} > \ln\frac{z}{z-2}.
\eeq
Therefore the ordered state occurs above a critical
temperature $T_c$ given by
\beq
\frac{V}{2kT_c} = \ln\frac{z}{z-2}.
\eeq

\subsection*{Peierls argument}

In the other paper, Peierls presents the proof that the Ising
model defined on the square lattice presents the ferromagnetic
state at sufficiently low temperatures. Peierls proof is as
follows. To each site of a square lattice, we assign a positive
or negative sign, representing the two states each site can
assume. The most common way to define the Ising model is to
designate the values $-J<0$ or $J>0$ for the energy between
two neighboring sites when they have the same or different
signs, respectively. However, it is more convenient to designate
the values 0 or $\varepsilon>0$, respectively, Thus, the total
energy is equal to the number of neighboring pairs of opposite
signs, multiplied by $\varepsilon$.

\begin{figure}
\begin{center}
\epsfig{file=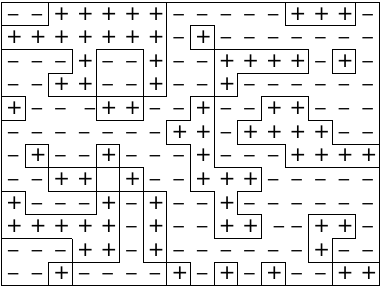,width=7.5cm}
\end{center}
\caption{Reproduction of the figure contained in Peierls paper
\cite{peierls1936} showing an example of boundary lines.}
\label{square}
\end{figure}

It is convenient to construct a line segment of length equal to
unity for each pair of neighboring sites with opposite signs,
as shown in figure \ref{square}.
A configuration of the entire lattice is composed of positive
and negative signs and lines that determine the boundary between
regions of opposite signs. Since each unit segment of the boundary
contributes an energy equal to $\varepsilon$, the total energy is
proportional to the total length of the boundary. It is important
to note that the boundaries completely determine the sign
configuration, except for the possibility of replacing all
positive signs with negative ones and vice versa, which is
unimportant for a zero field.

Peierls observes that there are closed and open boundaries.
The latter begin and end at the sides of the lattice. Peierls
demonstrates that at sufficiently low temperatures, the area
bounded by closed boundaries and intersected by open boundaries
is only a small fraction of the total area. Since subtracting
the area of regions of distinct signals is proportional do the
magnetization, this result shows that this quantity is strictly
nonzero, and therefore the model exhibits the ferromagnetic state.

Peierls demonstration, however, turned out to be defective but
amenable to correction. This was later done by Griffiths in 1964
\cite{griffiths1964}, who stated that Peierls argument
is valuable for explaining why the Ising model exhibits phase
transitions in two dimensions but not in a linear chain.
Furthermore, the argument can easily be extended to the
three-dimensional cubic lattice.

In his memoir \cite{peierls1985}, published in 1985, almost fifty
years after his demonstration, Peierls recalls the motivation that
led him to it. He says the inspiration came from his irritation
with certain works that seemed incorrect, unnecessarily complicated,
and based on unreasonable assumptions. This was the case with a
lecture on the Ising model given by a mathematician who argued that
the model would not exhibit the ferromagnetic state in two or three
dimensions. Peierls realized the inaccuracy, but instead of looking
for flaws in the argument, he sought to demonstrate that ferromagnetism
can occur in two and three dimensions. He then says he was surprised
to learn, a few years before 1985, that his 1936 argument had been
extensively used in statistical mechanics as one way to demonstrate
the existence of the ordered state.

\section{Conclusion}

We analyzed the emergence of the Ising model from its conception
by Lenz to Peierls, who unequivocally demonstrated that the model
possesses order in two dimensions. The success of the model is due
to the fact that it consists of only the necessary ingredients to
bring about the emergence of an ordered state. We saw how equivalent
models were introduced within the field of metalic alloys,
particularly by Bethe, who was concerned with demonstrating that
the model he analyzed does not sustain order in one dimension but
does possess order in two dimensions. This was subsequently
demonstrated by Peierls for the Ising model in a 1936 paper.
Peierls paper is also important for emphasizing that the models
used for metalic alloys by Bragg and Williams and by Bethe, as well
as the one used by Fowler for adsorption, are equivalent
to the Ising model.

The Ising model is usually considered a mathematical model, a model
that contains only the essential ingredients to obtain the properties
we wish to describe. In general, this term, mathematical model, is
used to denote theories that do not include all the factors that can
influence the construction of a more realistic theory. The usual
justification for introducing the mathematical model is that a
realistic theory would be too complicated both from the point of
view of construction and of obtaining results. However, we must keep
in mind that what is called a realistic theory may be devoid of
meaning, or at least needs to be better defined, since a scientific
theory is a construction, an invention of our minds.
The Ising model is a significant example of this human endeavor.

\section*{Acknowledgements}

I wish to thank Sílvio Salinas and Jürgen Stilck for their
suggestions and critical reading of the manuscript.


\end{document}